# Photo-excited Dynamics in the Excitonic Insulator Ta$_2$NiSe$_5$

Daniel Werdehausen[1,2], Tomohiro Takayama[1,3], Gelon Albrecht[1,2], Yangfan Lu[4], Hidenori Takagi[1,3,4], and Stefan Kaiser[1,2,*]

[1]Max-Planck-Institute for Solid State Research, 70569 Stuttgart, Germany

[2]4$^{th}$ Physics Institute, University of Stuttgart, 70569 Stuttgart, Germany

[3]Institute for Functional Matter and Quantum Technologies, University of Stuttgart, 70569 Stuttgart, Germany

[4] Department of Physics, The University of Tokyo, Bunkyo-ku, Tokyo 113-0033, Japan

*Corresponding author. Email: s.kaiser@fkf.mpg.de

**The excitonic insulator is an intriguing correlated electron phase formed of condensed excitons. A promising candidate is the small band gap semiconductor Ta$_2$NiSe$_5$. Here we investigate the quasiparticle and coherent phonon dynamics in Ta$_2$NiSe$_5$ in a time resolved pump probe experiment. Using the models originally developed by Kabanov et al. for superconductors [1], we show that the material's intrinsic gap can be described as almost temperature independent for temperatures up to about 250 K to 275 K. This behavior supports the existence of the excitonic insulator state in Ta$_2$NiSe$_5$. The onset of an additional temperature dependent component to the gap above these temperatures suggests that the material is located in the BEC-BCS crossover regime. Furthermore, we show that this state is very stable against strong photoexcitation, which reveals that the free charge carriers are unable to effectively screen the attractive Coulomb interaction between electrons and holes, likely due to the quasi one-dimensional structure of Ta$_2$NiSe$_5$.**

# 1. Introduction

The excitonic insulator state can be formed in both a semiconductor that exhibits a small band gap or a semimetal which is distinguished by a small band overlap [2-8]. In the semiconducting case, the system is unstable against the formation of excitons if the exciton binding energy is larger than the system's semiconducting band gap. The consequence is the formation of a high number of excitons, which leads to a characteristic broadening of the energy gap (left side of figure 1a) [8]. At high exciton densities, i.e. if the exciton wave functions overlap significantly, these "preformed" excitons can, in a BEC-like transition, condense into the coherent excitonic insulator ground state [8]. In contrast, in the semimetallic regime, the transition into the excitonic insulator state takes place if the charge carrier density becomes so low that the screening of the Coulomb interaction between electrons and holes breaks down (right side of figure 1a). In this case the point at which the excitons are formed coincides with the phase transition, and the latter can be described in analogy to BCS-theory [7,8]. These considerations are summarized in the phase diagram shown in figure 1b [8]. In both regimes the excitonic insulator ground state is consequently distinguished by a characteristic band flattening (see figure 1a) and the size of the gap around the Fermi level is determined by the binding energy of the excitons [8]. The properties of this gap can, for example, be probed using ARPES or infrared measurements of the optical conductivity [9-12]. However, it has also been shown for high temperature superconductors [1,13-15], and other materials [16-18] that time resolved spectroscopy can provide additional insight into the energy structure around the Fermi level: On the one hand the quasiparticle dynamics allow to extract information about the properties of the gap [1,13,14,16-18], and on the other hand the dephasing times of coherent phonons that are excited in time resolved experiments depend critically on the available decay channels, which in turn are influenced by the nature of the electronic gap [15]. Here we use this approach to investigate the energy structure of $Ta_2NiSe_5$. $Ta_2NiSe_5$ has recently become one of the most promising candidates for a purely electronically driven excitonic insulator [9,10,12,19-22]. As shown in figure 1c, the structure of $Ta_2NiSe_5$ consists of one dimensional Ni- and Ta-chains which run along the a-axis of the crystal [20]. The material's electronic structure around the Fermi level is also quasi one dimensional, in the sense the Ni chains supply the valance band and the Ta-chains provide the double degenerate conducting bands. Excitons in the system are consequently formed as charge transfer excitons between a hole on a Ni-chain and an electron on a Ta-chain [20]. This spatial separation appears to prevent an efficient screening of the attractive Coulomb interaction and is therefore responsible for the high exciton binding energies. Furthermore, the spatial separation may also lengthen the exciton life times. Moreover, band structure calculations revealed that without taking the formation of excitons into account, $Ta_2NiSe_5$ is a direct semiconductor with a very small band gap [20]. All these findings make $Ta_2NiSe_5$ an ideal candidate for the realization of the excitonic insulator ground state. In fact, recent resistivity and thermodynamic measurements revealed that indeed the existence of a new phase below $T_c$=328 K [12], for which ARPES measurements showed that the bands are characteristically flattened opening a gap at low temperatures [9,10,21]. In addition, distinct exciton resonances were found in optical conductivity measurements [19], which show that self-localized excitons are indeed responsible for the opening of the electronic gap. However the existence of a coherent condensate of excitons forming an excitonic insulator phase is still under debate. Recent time dependent measurements under high excitation fluence found a low frequency phonon showing amplitude mode like behavior revealing a strong fingerprint towards the existence of a coherent excitonic insulator ground state [22].

## 2. Experimental setup and results

In order to further elucidate the electronic structure and show that the quasiparticle relaxation dynamics support the assumption of a BEC-like transition into the excitonic insulator state at $T_c$=328 K, we performed pump-probe experiments using 1.55 eV (800 nm), 130 fs laser pulses. Probing the system's relaxation dynamics and its coherent oscillations in temperature and excitation density dependent measurements below and above the transition temperature we aim to test up to what extend the scenario of an temperature independent gap of a BEC holds for the excitonic insulator. In this the gap size should be solely be defined by the exciton binding energy of the preformed excitons above Tc. However we note that optical probes do show a gradual fill-in of the gap on increasing temperature [12].

Figure 2a) depicts the time-dependence of the photoinduced reflectivity changes ΔR/R(t) at perpendicular polarization to the chains at different temperatures[i]. The measured time-trace is a superposition of the electronic signal and several coherent oscillations. As discussed in detail in the following, the electronic part describes the quasiparticle excitation dynamics and their subsequent recombination across the gap. To analyze the signal's electronic component and extract the coherent oscillations a suitable function was fitted to the data:

$$\frac{\Delta R(t)}{R} = \left( A_1 \exp\left(-\frac{t-t_0}{\tau_1}\right) + A_2 \exp\left(-\frac{t-t_0}{\tau_2}\right) \right) \times \left( \text{erf}\left(\frac{t-t_0}{\Delta T_{O1}}\right) + 1 \right) + \left( \text{erf}\left(\frac{t-t_0}{\Delta T_{O2}}\right) + 1 \right) \times A_\infty. \quad (1)$$

---

[i] Recently we already showed that the coupling to the exciton condensate is very high at parallel polarization to the Ni- and Ta-chains, which leads to a complete depletion of the condensate even at low excitation densities [22].

This function is motivated as follows: The two error functions account for the two different rise times ($T_{O1}$ and $T_{O2}$) of the onset. Fits to multiple sets of data showed that the fast onset component (rise time $T_{O1} \approx 100$ fs) emerges into a subsequent double exponential decay (relaxation times $\tau_1$ and $\tau_2$), whereas the slow component (rise time $T_{O2} \approx 1$ ps) describes the onset of the constant value ($A_\infty$) the signal converges to at large time delays. Throughout this paper this fit ansatz is used to determine the time constants ($T_{O1}, T_{O2}, \tau_1$ and $\tau_2$), as well as the amplitudes ($A_1$, $A_2$ and $A_\infty$) of the electronic signal. Besides providing the basis for the analysis of the electronic signal, the fit ansatz also allows extracting the coherent oscillations by subtracting the fit from the measured signal. This is exemplarily shown in the inset to figure 2b) on the measurement at 80 K. The corresponding FFTs at 80 K and 350 K are presented in the main panel of figure 2b). It can be seen that at 80K the spectrum contains three distinct modes at 1,01 THz, 2,07 THz and 2,98 THz (from here on denoted as 1 THz, 2 THz and 3 THz modes). Using LDA calculations and Raman measurements the 1 THz oscillation was identified as an $A_{1g}$ mode, whereas the 2 THz (3 THz) mode is a $B_{1g}$ ($A_{1g}$) phonon [22]. To investigate the behavior of the modes quantitatively, we applied FFT band pass filters with a bandwidth of 1 THz around the respective peaks. This is shown in figure 2c) on the example of the 3 THz mode. We fitted a damped harmonic oscillator function to the extracted oscillations to determine the modes' initial amplitudes, frequencies and dephasing times. A detailed analysis of the temperature and excitation density dependence of the modes' initial amplitude and frequency was already given elsewhere [22]. In particular the 1 THz mode at high excitation densities is shown to couple in a non-trivial way to the excitonic insulator ground state and shows an amplitude mode like behavior. Here we will instead focus on the 3 THz coherent phonon mode's dephasing time for probing existing decay channels.

The photon energy of 1.55 eV utilized in our experiment is much larger than the size of the optical gap of Ta$_2$NiSe$_5$ (2$\Delta \approx$160-180 meV at low temperatures). This makes it impossible to directly identify the dominant excitation and relaxation mechanisms in the band structure. However, in their comprehensive work on high temperature superconductors Demsar, Kabanov et al. have shown that the dynamics observed in an 800nm pump probe experiment in these materials nevertheless depend critically on the properties of the energy gap around the Fermi level [1,13]. Their models allow to determine the gap's properties from temperature dependent measurements of the photoinduced reflectivity changes, and were found to provide a good agreement with the experimental results [1,13,14,16-18]. Here we show that these models also us to describe dynamics observed in Ta$_2$NiSe$_5$: The results of our analysis support the assumption of a BEC-like phase transition into the excitonic insulator ground state, but also show that there is a an additional temperature depended contribution to the gap.

### 3. Analysis and Discussion

In general, it can be assumed that in all these materials the absorption of a pump photon leads to the generation of an electron hole pair [23]. Since the energy of a pump photon is large compared to the size of the gap, the photoexcited quasiparticles possess a large amount of excess kinetic energy. In the initial state, directly after the excitation the distribution of the excited quasiparticles is consequently nonthermal, in the sense that their distribution can't be described by a Fermi-Dirac function with an effective temperature. The generally accepted model is that, in the first relaxation step these highly excited quasiparticles now thermalize among themselves through electron-electron scattering. In this process each quasiparticle can excite multiple other electron hole pairs [23]. This leads to a quasiparticle avalanche, in which each initially excited particle creates a high number of secondary excitations. Kabanov et al. assumed that the gap in the density of states around the Fermi level now represents a bottleneck, which prevents the excited quasiparticles from reaching the states below the gap [1]. The quasiparticles consequently accumulate in states above the gap on a time scale of $\tau_{ac} \approx$ 10-100 fs [23]. In this state the high number of quasiparticles accumulated above the gap consequently form a quasi-equilibrium at an effective temperature. This sea of particles was assumed to give rise to the observed change in the macroscopic reflectivity ($\Delta$R(t)/R) [1]. The initial onset of the signal in $\Delta$R(t)/R can therefore be attributed to the quasiparticle avalanche. Consequently, the time scale of the subsequent decay of $\Delta$R(t)/R is characterized by the time it takes the quasiparticles to relax back across the gap [1].

*3.1 Photo-Excited Carrier Amplitudes*

Assuming that the amplitude of the reflectivity changes (A(T)) is proportional to the number of particles that are accumulated above the gap, one can derive the temperature dependence of the amplitude A(T). In the case of a temperature independent gap ($\Delta_{EI}$ for the excitonic gap) this relationship reads [1]:

$$A(T) = \frac{A/\Delta_{EI}}{1+B\,\exp\left(-\frac{\Delta_{EI}}{k_B T}\right)}, \qquad (2)$$

where A and B are treated as free fit parameters in the following. Such a temperature independent gap is expected for a BEC-like phase transition, since in this case the particles are formed above the critical temperature and only condense into a coherent ground state on going through the phase boundary [1]. Both above and below the critical temperature the size of the gap is therefore denoted by the exciton binding energy.

Figure 3 presents the temperature dependence of the amplitudes $A_1$, $A_2$ and $A_\infty$ (see equation 1). A comparison of $A_1$ (figure 3a) and $A_\infty$ (figure 3c) shows that both amplitudes exhibit a qualitatively very similar behavior. Both amplitudes are almost constant at low temperatures, but decrease linearly at high temperatures. This is exactly the behavior that is predicted by equation 2 for a temperature independent gap. Fitting equation 2 to the temperature dependence of the amplitude $A_1$ (figure 3a) (neglecting the data point at 80 K) yields a value of $\Delta_{EI}$ = 103 ± 20 meV. This suggests that this component of the double exponential decay (amplitude $A_1$ and time constant $\tau_1$) indeed describes the recombination of quasiparticles into excitons across a temperature independent gap. As discussed in the following, this assumption is also confirmed by the temperature dependence of the time constant $\tau_1$ and the excitation density dependence of the amplitude $A_1$. The deviation of the data point at 80 K from the expected behavior can probably be attributed to pump induced changes to the gap: The fit ansatz in equation 2 only holds true in a weak perturbation regime, in which the pump doesn't induce any changes to the gap. But, recent time resolved ARPES measurements on $Ta_2NiSe_5$ showed that, depending on the excitation density, the pump can either induce a narrowing or a broadening of the gap [24]. Possibly these changes are also strongly temperature dependent and could explain the deviation of the data from the fit. As discussed below this is confirmed by the observation that the temperature dependence of the amplitude $A_\infty$ doesn't exhibit a similar discrepancy. To this end Figure 3c depicts the amplitude $A_\infty$ as a function of temperature. As already mentioned in the discussion of equation 1, the amplitude $A_\infty$ characterizes the constant value the signal converges to at large time delays. Measurements over a larger time delay range ($\approx$ 150 ps) showed that this component has a life time that exceeds the accessible time window by several orders of magnitude. The long lifetime and the plateau's particularly slow onset ($T_{O2} \approx$ 1 ps) indicate that this component is caused by bolometric heating. In fact, it is well known that the propagation of heat out of the pumped area can take tens to hundreds of nanoseconds [17,18]. This is confirmed by the finding that the behavior of $A_\infty$ is well reproduced by equation 2. The reason for this is that the relationship for a temperature independent gap (equation 2) also holds true if all degrees of freedom in the pumped area have already reached quasi-equilibrium, but the pumped spot itself still has a significantly higher effective temperature than its surroundings. Further evidence that the long lived component is indeed caused by bolometric heating is given in the discussion of the excitation density dependence of the amplitude $A_\infty$. Fitting equation 2 to the temperature dependence of $A_\infty$ yielded a value of $\Delta_{EI}$ = 102 $\pm$ 16 meV, which agrees perfectly with the value obtained from the fit to the amplitude $A_1$. Furthermore, the fact that $A_\infty$ agrees much better with the fit in the low temperature regime than $A_1$, underpins the assumption that the photoinduced changes of the gap are indeed responsible for the deviation of $A_1$ from the expected behavior in this temperature range: The time resolved ARPES measurements showed that the recovery time of the gap is in the order of 1 ps [24]. At large time delays the gap has consequently already recovered and the deviation from the fit is therefore not present in the temperature dependence of the amplitude $A_\infty$. The value of $\Delta_{EI} \approx$ 100 meV obtained from the fits to both $A_1$ and $A_\infty$ agrees roughly with the value of $2\Delta \approx$ 160-180 meV determined from ARPES and optical conductivity measurements at low temperatures [12,21] and an approximate transport gap of about 100 meV [19]. On the one hand the discrepancies can possibly be attributed to the fact that the time resolved measurements are probing an excited state and not the ground state itself: In this excited state a significant number of excitons are heated out of the ground state. Therefore on the one hand the value of $\Delta_{EI}$ obtained from the fits could therefore describe an "average binding energy" of these hot excitons. On the other hand this could also be an indication of a non fully BEC nature of the gap. Both amplitudes ($A_1$ and $A_\infty$) agree well with the fits for low temperatures but they show a small but distinct drop between 250 K and 275 K (indicated by the arrows in Figs. 3(a) and (c)). Such a drop is characteristic of a temperature dependent gap (BCS-like gap), where a sharp decrease of the photoinduced reflectivity changes occurs in the vicinity of the critical temperature and the signal vanishes when the gap is completely closed [1]. The sharp drops in $A_1$ and $A_\infty$ therefore indicate a partial closing of the electronic gap, in agreement with the partial fill-in of the gap probed by optics [12]. As discussed below this is also confirmed by the temperature dependence of the time constant $\tau_1$ and the 3 THz mode's dephasing time (Fig. 4(a)), which also exhibit distinct anomalies in the temperature regime between 250 K and 275 K. This suggests that the gap also has a small additional BCS-like contribution, and is consequently not completely temperature independent. Such a behavior is actually expected for $Ta_2NiSe_5$, since pervious measurements showed that the material sits close to the $\Delta E(T)=0$ line in the phase diagram (figure 1a) [19]. $Ta_2NiSe_5$ is consequently located in the BEC to BCS crossover regime, where a shift from a BEC-like condensation (semiconducting limit) to a BCS-like behavior (semimetallic limit) takes place.

To conclude the discussion of the amplitudes $A_1$, $A_2$ and $A_\infty$, figure 3c presents the temperature dependence of the amplitude $A_2$ (see discussion of equation 1). It can be seen from the figure that this amplitude doesn't show a systematic temperature dependence. As indicated by the linear fit to the data the amplitude seems to be almost temperature independent. However, the general features, in particular the peak at 160 K, were present in all analyzed data sets. But since there is no distinct trend and the data doesn't agree with the behavior commonly observed in other materials [1,13,14,16-18], a direct identification of the microscopic process responsible for this component on the basis of an 800 nm pump probe experiment only is not possible.

*3.2 Carrier Relaxation Dynamics*

Besides the electronic amplitudes, the respective relaxation constants can provide further insight into the nature of the gap. Figure 4 depicts $\tau_1$ and $\tau_2$ as a function of temperature. From the figure's left panel it can be seen that the time constant $\tau_1$ remains almost constant at low temperatures, then decreases sharply around 250K, reaches a minimum at 275K, and finally increases linearly in the high temperature regime. The behavior in the regime below the minimum at 275K is well reproduced by the fit ansatz [14]:

$$\tau(T) = \frac{1}{A + B\sqrt{\Delta T} \exp\left(-\frac{\Delta_{EI}}{k_B T}\right)}, \qquad (3)$$

where A and B were again treated as free parameters, and $\Delta_{EI}$ denotes the size of the gap. Equation 3 was derived for superconductors and holds true at temperatures that are small compared to $T_c$. Fitting equation 3 to the data yields $\Delta_{EI}$ = 96 ± 25 meV. This value again agrees perfectly with the results from the previous fits to the amplitudes $A_1$ and $A_\infty$. The good agreement underpins the assumption that the relaxation component characterized by $A_1$ and $\tau_1$ represents the recombination of particles into excitons. Moreover, since the relaxation time is expected to be very sensitive to changes of the gap, the sharp turn at 275K shows that the gap is indeed changing significantly around this temperature. This can be attributed to the previously discussed BCS-like closing of the gap. Finally, figure 4c depicts the temperature dependence of the relaxation time $\tau_2$. Within the error bars it remains constant at both low and high temperatures, but a distinct step between 200 K and 225 K is observed.

As discussed in the introduction the electronic gap also critically affects the dephasing times of the coherent phonons. To conclude the discussion of the temperature dependence figure 5 therefore depicts the 3 THz phonon's dephasing time as a function of temperature at two different excitation densities. At both excitation densities the dephasing time exhibits qualitatively the same behaviour: At low temperatures the dephasing time decreases linearly with temperature. Around 250 K, however, the slope of the linear relationship changes abruptly, and on further increasing the temperature the dephasing time again decreases linearly, but with a much smaller slope. A similar behaviour was also observed in superconductors [15]. In this class of materials, the kink in the dephasing time occurs in the vicinity of the $T_c$, which was attributed to the fact that the emergence of the gap in the electronic structure significantly changes the available decay channels [15]. The sharp kink in the 3 THz phonon's dephasing time therefore confirms that there is a partial onset of an additional contribution to the gap, possibly a BCS-like closing of the electronic gap that sets in above 250 K to 275 K.

*3.3 Excitation-Fluence Dependence*

All experimental results discussed until now can be well explained by the assumption that $Ta_2NiSe_5$ is located in the BEC to BCS crossover regime of the excitonic insulator phase: In our measurements the material's gap appears to be almost temperature independent, but it exhibits a small, but distinct additional component, likely a BCS-like closing at higher temperatures. To gain further insight into the dynamics in the material we also varied the excitation density at a fixed temperature. This allows to investigate the change of the dynamics on increasing the excitation density from close to equilibrium conditions to a strong excitation regime. We used excitation densities of up to 1.6 mJ/cm$^2$; higher excitation densities irreversibly damaged the samples.

In order to show that the signal's initial steep onset is indeed caused by a quasiparticle avalanche, figure 6 presents the excitation density dependence of the rise times $T_{O1}$ and $T_{O2}$ (see equation 1 and subsequent discussion). It can be seen from the figure's left panel that the faster rise time ($T_{O1}$) decreases with power and eventually runs into saturation. The asymptotic behavior at high excitation densities is due to the limited time resolution of our experiment. The pronounced decrease, on the other hand, is in line with the assumption that the fast increase is caused by a quasiparticle avalanche [18]. The slower rise time ($T_{O2}$; right panel of figure 6), in contrast, remains constant within the errors bars across the entire excitation density range. This is in line with the assumption that this component can be attributed to a bolometric response.

Figure 7 presents the electronic amplitudes $A_1$, $A_2$ and $A_\infty$ as a function of power. It can be seen from the figure's upper left panel that the amplitude $A_1$ exhibits a distinct saturation behavior at high excitation densities. Such a trend is expected for any excitation from a finite electron reservoir, and was also observed in other materials [18,23]. This confirms that this component characterizes the number of quasiparticles that are accumulated above the gap. The slower component's amplitude ($A_2$), in contrast, increases linearly up to a distinct threshold (0.9 mJ/cm$^2$), after which it saturates abruptly (figure 7b). As mentioned in the discussion of the temperature dependence the physical process behind this component unfortunately can't be identified on the basis of an 800 nm pump probe measurement alone. To conclude the discussion of the electronic amplitudes, the excitation density dependence of the long lived component's amplitude $A_\infty$ is depicted in figure 7c. It can be clearly seen that this amplitude increases linearly with power. Since the temperature of the pumped area after the excitation is expected to increase linearly with pump power, this linear relationship confirms that the long lived component is indeed caused by a bolometric response. However, this also indicates that the amount of energy that is absorbed in the pumped area increases linearly with power, and consequently that the percentage of pump photons that is absorbed remains constant. This suggests that the asymptotic behavior in the amplitude $A_1$ is not caused by a saturation of the number of photons that are absorbed, but rather by a saturation of the quasiparticle avalanche.

To conclude the discussion of the excitation density dependence, figure 8 presents the relaxation times $\tau_1$ and $\tau_2$ as a function of power. The figure's upper left panel shows that the fast component ($\tau_1$) initially decreases at small excitation densities. But on further increasing the power it starts to increase and finally saturates at high excitation densities. This trend coincides with the behavior of the gap that was found in the recent time-resolved ARPES measurements [24]: These measurements have shown that at low excitation densities the pump pulses induce a narrowing of the gap. For excitation densities exceeding 0.2 mJ/cm$^2$, however, the gap was found to increase directly after the excitation before heating effects set in. Since the lifetime $\tau_1$ is expected to depend strongly on the size of the gap (see for example equation 3), this could explain its non-monotonic behavior. For the sake of completeness figure 8b) depicts the relaxation time $\tau_2$ as a function of power. This component appears to increase slightly with power, but the error bars are too large to obtain reliable conclusions.

Summarizing the investigation of the excitation density dependence of the electronic signal of $Ta_2NiSe_5$ shows that the amplitude of the electronic signal saturates at high excitation densities, which indicates that the pump pulses strongly deplete the electronic reservoir in this regime. The relaxation times $\tau_1$ and $\tau_2$, however, are only weakly power dependent, even at high excitation densities. This contrasts with the behavior observed in superconductors, where the lifetimes depend strongly on the excitation density in the weak perturbation regime, and eventually reach saturation at high pump powers [23]. In this class of materials this behavior was attributed to a photo-induced phase transition, in which the superconducting phase is completely destroyed [23]. The lack of such a strong excitation density dependence in $Ta_2NiSe_5$ suggests that no phase transition is induced, and that the exciton condensate consequently remains intact. This is confirmed by the time-resolved ARPES measurements, where, as already mentioned, the size of the gap was found to even increase at high excitation densities [24]. This proofs that the excitonic insulator state is not destroyed, even at high excitation densities. This finding is very surprising and contrasts with the behavior that was found in other potential excitonic insulators: Time resolved ARPES measurements on $1T$-$TiSe_2$ showed that the electronic order in this material breaks down at intermediate excitation densities [25]. This photoinduced phase transition was assumed to be caused by the photoexcited quasiparticles, which screen the attractive Coulomb forces between electrons and holes in the transient state, and thus destroy the gap [25]. This raises the question of why the photoexcited quasiarticles in $Ta_2NiSe_5$ are unable to effectively screen the Coulomb interaction. The answer to this question could help to understand why the exciton states in the materials are so remarkable stable that they can form the excitonic insulator phase even above room temperature. Intuitively one can assume that this can be attributed to the quasi one-dimensional structure, which, as discussed manifests itself in the formation of self localized or charge transfer excitons between the chains that do not allow forming a quasi-metallic transport under photoexcitation.

**4. Summary and Conclusion**

In conclusion, we have shown that 800 nm pump probe measurements on $Ta_2NiSe_5$, allow to extract information about the nature of the excitonic gap. We found the gap to be almost temperature independent, with a small temperature dependent contribution that closes between 250 K and 275 K. Therefore our results support the assumption that $Ta_2NiSe_5$ is indeed an excitonic insulator, and that the material is located in the BEC-BCS crossover regime. Furthermore, our measurements revealed that the excitonic insulator state remains intact after photoexcitation, even at excitation densities close to the damage threshold. This shows that the quasiparticles are unable to effectively screen the attractive Coulomb interaction between electrons and holes. Determining the microscopic origin of this inefficiency will be a key task in future studies, since the understanding of this peculiarity would provide direct insight into the exciton and quasiparticle states in the material. Finally, the fact that the time-resolved measurements allow to clearly distinguish a temperature-independent gap from a BCS-like gap makes the relaxation dynamics and ideal probe to investigate BEC-BCS crossover that can be induced in the material class $Ta_2Ni(Se_{1-x}S_x)_5$: Starting from $Ta_2NiSe_5$ the gap can be tuned towards the semimetallic side (BCS regime) by applying physical pressure, and towards the semiconducting side by increasing the sulphide concentration ($0<x\leq1$) [19]. Investigating the relaxation dynamics at various different points would therefore allow to map out the properties of the gap across the entire phase diagram of the excitonic insulator phase (figure 1a).

**Acknowledgements:**

The authors thank T. Larkin, A.V. Boris, and A. Yaresko for fruitful discussions and sharing their insights from optical measurements and band structure calculations. We also acknowledge support by the BW Stiftung and the MWK Baden-Württemberg through the Juniorprofessuren-Programm as well as the Daimler und Benz Stiftung for their support.

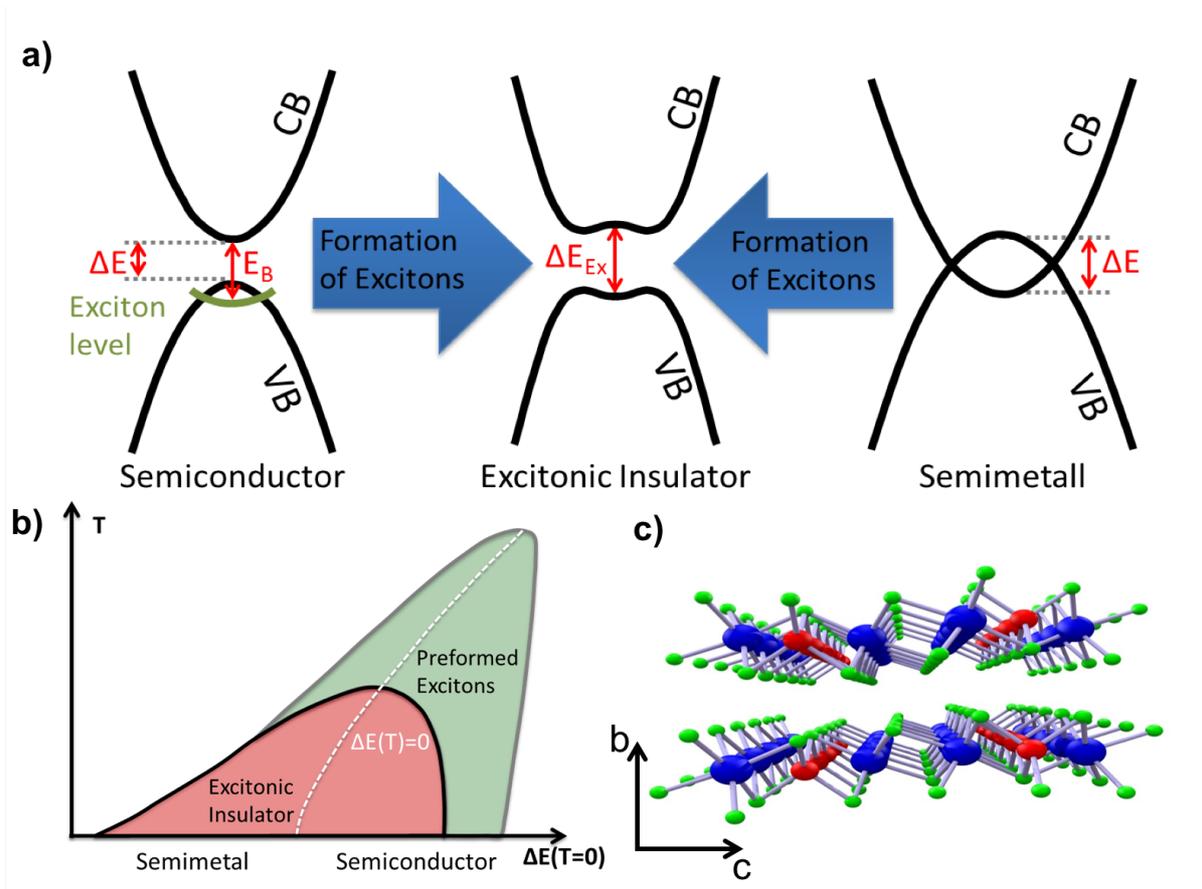

*Figure 1: (a) Transition into the excitonic insulator state: In a semiconductor a characteristic band flattening occurs if the exciton binding energy exceeds the band gap, whereas in a semimetall the transition takes place if the screening of the Coulomb interaction breaks down. (b) Phase diagram: Red denotes the excitonic insulator phase and green the regime in which a high number of excitons are formed, but no condensation takes place. The dashed white line separates the semimetallic from the semiconducting phase without the formation of excitons. (c) Structure of $Ta_2NiSe_5$: Red represents Ni, blue Ta and green Se. The material is made up of Ni and Ta chains, which run along the a axis of the crystal.*

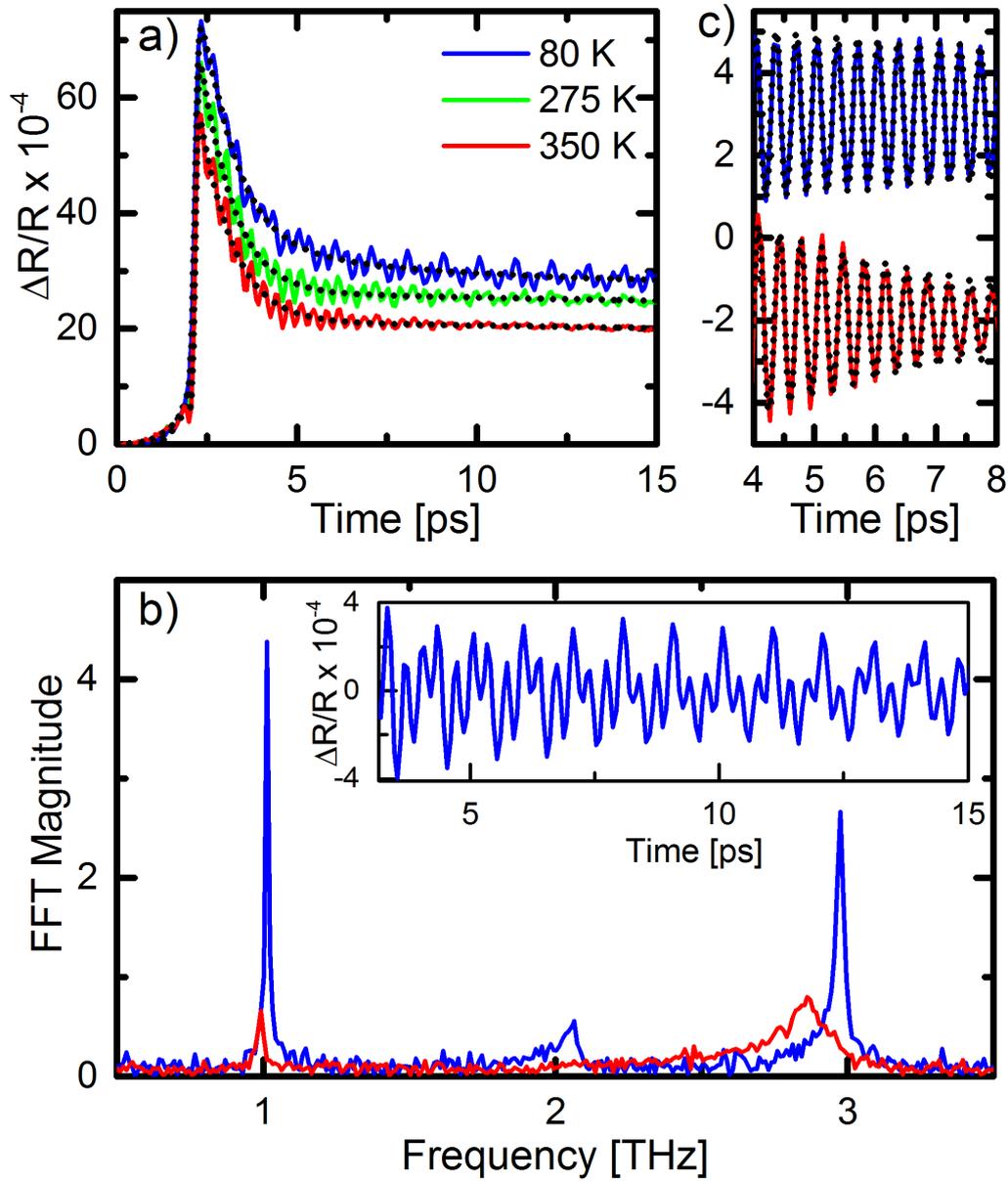

*Figure 2: Time-trace of photoinduced reflectivity changes at different temperatures and an excitation density of 0.35mJ/cm². The signal is made up of the electronic response and the coherent oscillations. The dotted black lines represent fits to the measured data. The inset of (b) shows only the coherent oscillations, which were extracted by substracting the fits in (a) from the measured data. The main panel of (b) presents the corresponding FFTs at 80K and 350K. (c) depicts the coherent phonon mode at 3 THz, which was extracted using a FFT band pass filter. The amplitude, frequency and dephasing time of the phonon mode was determined by fitting a damped harmonic oscillator to the data (dotted lines).*

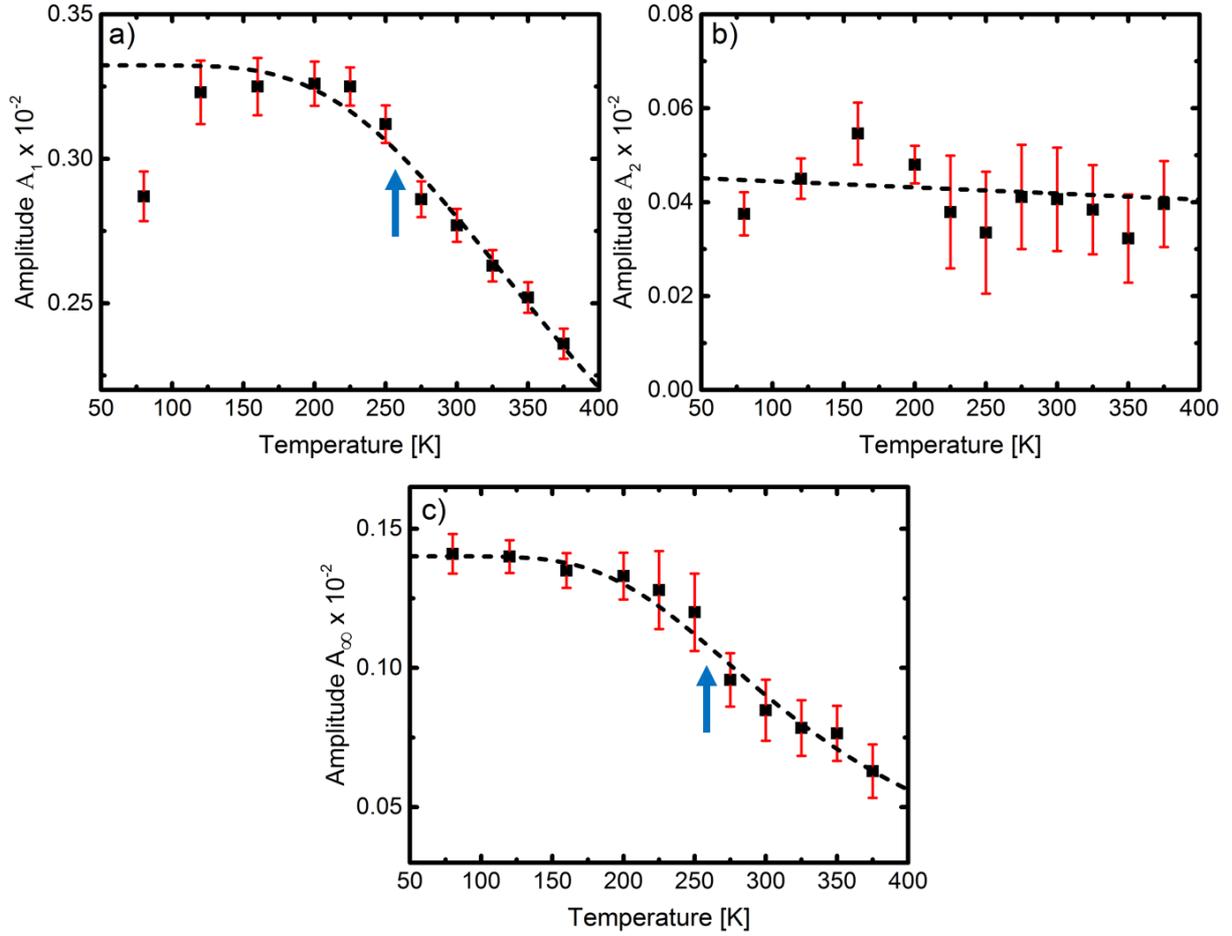

*Figure 3: Temperature dependence of the electronic amplitudes defined in the fit ansatz in equation 1 at an excitation density of 0.35 mJ/cm$^2$. $A_1$ (a) denotes the amplitude of the component with the shortest lifetime ($\tau_1 < 1$ ps), $A_2$ (b) represents the amplitude of the component with the intermediate lifetime ($\tau_2 \approx 10$ ps), and $A_\infty$ is the amplitude of the long lived component. The function that was fitted to $A_1$ and $A_2$ holds true for a temperature independent gap (see discussion of equation 2). The blue arrows in a) and b) indicate the potential onset of a partial BCS-like closing of the electronic gap.*

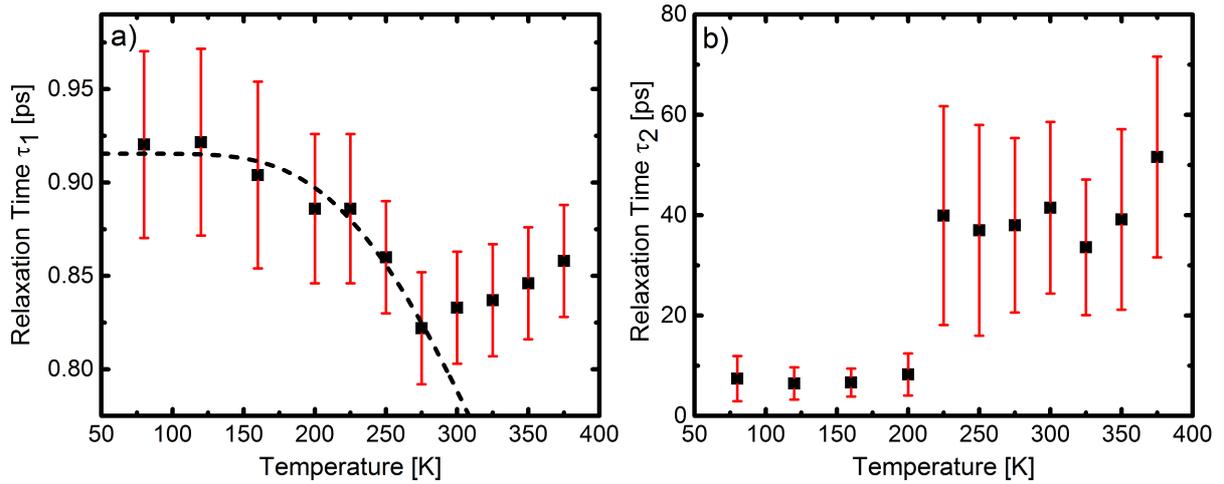

*Figure 4: Temperature dependence of the relaxation times $\tau_1$ (a) and $\tau_2$ (b) at an excitation density of 0.35 mJ/cm². For the fit in (a) only points up to 275K were used. The distinct kink at 275 K is likely due to the partial BCS-like closing of the electronic gap. At temperatures above 200 K the lifetime $\tau_2$ (b) exceeds the measured time window, explains the large errorbars obtained from the fits.*

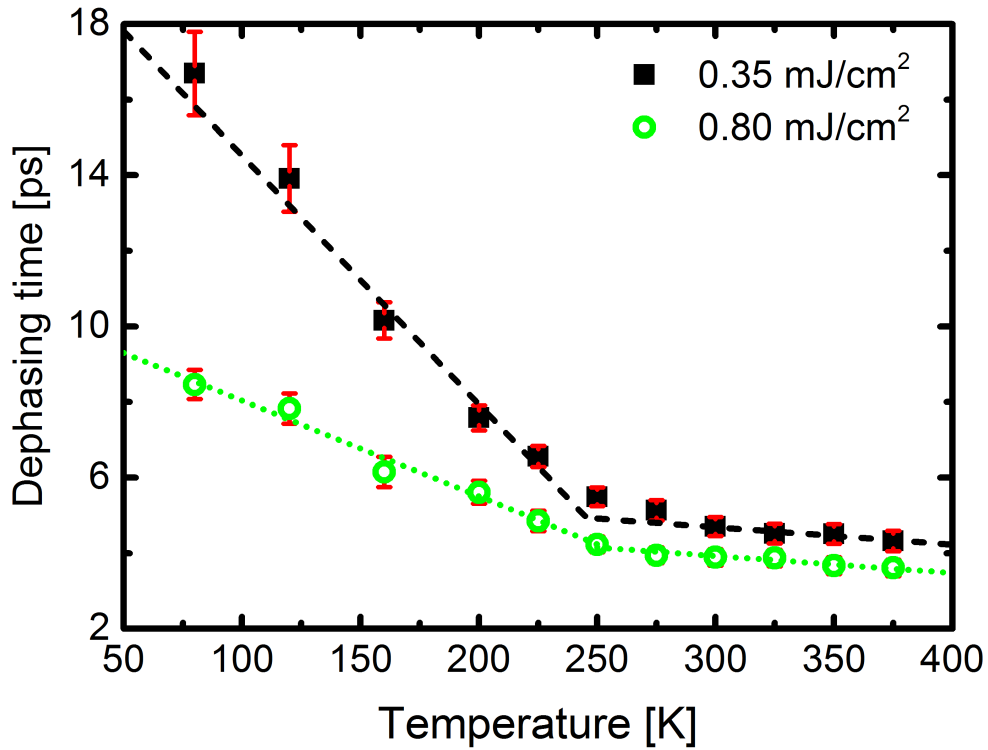

*Figure 5: Temperature dependence of the 3 THz mode's dephasing time at two excitation densities (0.35 mJ/cm² and 0.80 mJ/cm²). The kink at 250 K can be attributed to the partial BCS-like closing of the electronic gap.*

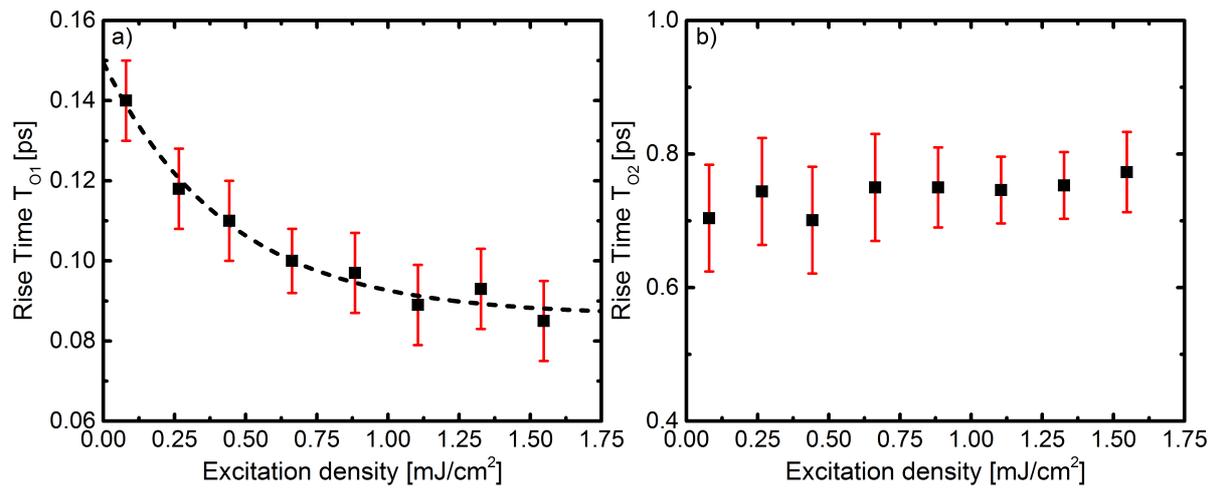

*Figure 6: Excitation density dependence of the rise times $T_{O1}$ (a) and $T_{O2}$ (b) at 120K. The pronounced decrease of $T_{O1}$ (a) is characteristic for a quasiparticle avalanche and the saturation at high excitation densities can be attributed the limited time resolution.*

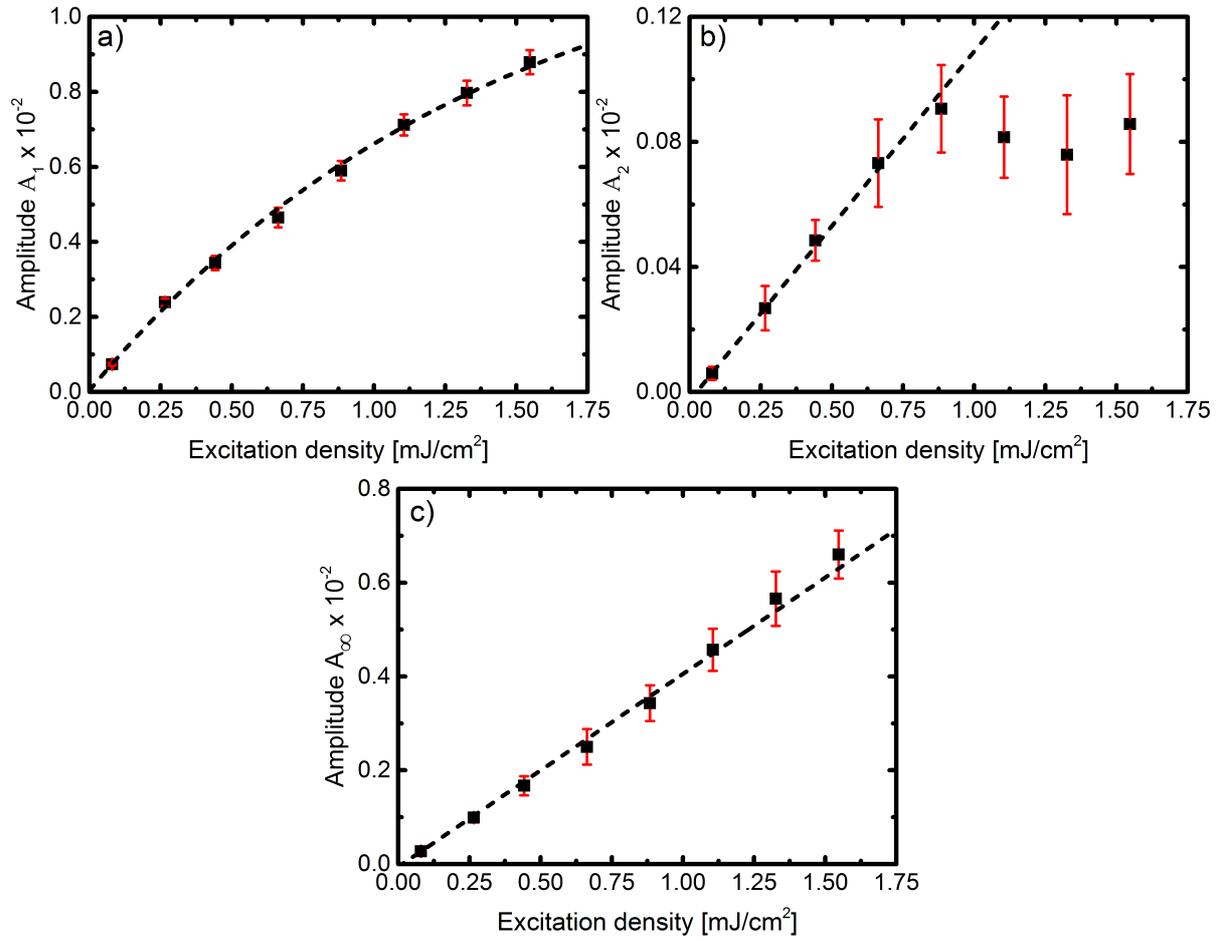

*Figure 7: Excitation density dependence of the electronic amplitudes defined in the fit ansatz in equation 1 at a temperature of 120 K. $A_1$ (a) denotes the amplitude of the component with the shortest lifetime ($\tau_1 < 1$ ps), $A_2$ (b) represents the amplitude of the component with the intermediate lifetime ($\tau_2 \approx 10$ ps), and $A_\infty$ is the amplitude of the long lived component. In (a) the function $A_1(\rho)=a_0(1-\exp(\rho/\rho_c))$ was fitted to the data to emphasize the saturation behavior. In a) and b) linear functions were fitted to the data as guides to the eye (in b) only data points up to 0.9 mJ/cm² were used for the fit).*

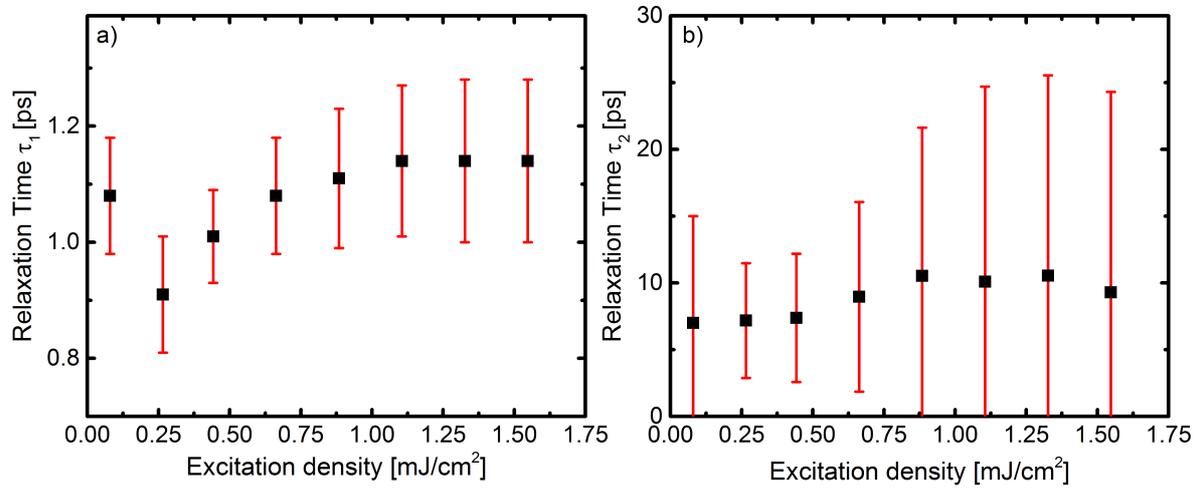

*Figure 8: Excitation density dependence of the relaxation times τ₁ (a) and τ₂ (b) at a temperature of 120 K. For the fit in (a) only points up to 275K were used. The trend of the relaxation time τ₁ (a) coincides with the behavior of the gap found in time resolved ARPES measurements.*